\documentclass[preprint2]{aastex}

\newcommand\br{\mbox{$B\!-\!R$}}
\newcommand\jh{\mbox{$J\!-\!H$}}
\newcommand\hk{\mbox{$H\!-\!K_s$}}
\newcommand\vh{\mbox{$V\!-\!H$}}

\shorttitle{CG~J1720-67.8: TDG candidates}

\received{2002 February 7}
\begin{document}

\title{Candidate Tidal Dwarf Galaxies in the Compact Group CG~J1720-67.8\footnote{Partly 
based on data collected at the European Southern Observatory, La Silla, Chile (Proposal
63.N-0737) and at the Anglo Australian Telescope, Siding Spring, Australia (Proposal PATT/02A/24).}}

\author{S. Temporin and R. Weinberger}
\affil{Institut f\"ur Astrophysik, Leopold-Franzens-Universit\"at Innsbruck, 
Technikerstra\ss e 25, A-6020 Innsbruck, Austria}
\email{giovanna.temporin@uibk.ac.at, ronald.weinberger@uibk.ac.at}

\author{G. Galaz}
\affil{Departamento de Astronom\'{\i}a y Astrof\'{\i}sica, Pontificia Universidad Cat\'olica de Chile,
Casilla 306, Santiago 22, Chile}
\email{ggalaz@astro.puc.cl}

\and
\author{F. Kerber}
\affil{Space Telescope European Coordinating Facility, European Southern 
Observatory, Karl-Schwarzschild-Stra\ss e 2, D-85748 Garching, Germany}
\email{fkerber@eso.org}

\begin{abstract}
This is the second part of a detailed study of the ultracompact group CG~J1720-67.8:
in the first part we have focused the attention on the three main galaxies of the group
and we have identified a number of candidate tidal dwarf galaxies (TDGs).
Here we concentrate on these candidate TDGs.
Absolute photometry of these objects in $BVRJHK_s$ bands confirms 
their relatively blue colors, as we already expected from the inspection of optical 
and near-infrared color maps and from the presence of 
emission-lines in the optical spectra. The physical conditions in such candidate TDGs are 
investigated through the application of photoionization models, while the optical colors are 
compared with grids of spectrophotometric evolutionary synthesis models from the literature.
Although from our data self-gravitation cannot be proved for these objects, their general 
properties are consistent with those of other TDG candidates. 
Additionally we present the photometry of a few ``knots'' detected in the immediate 
surroundings of CG J1720-67.8 and consider the possibility that these objects might
belong to a dwarf population associated with the compact group. 
\end{abstract}

\keywords{galaxies: evolution --- galaxies: interactions --- galaxies: 
starburst}

\section{INTRODUCTION}
An often neglected aspect of interaction-induced galaxy evolution is the possibility of 
formation of dwarf galaxies out of interaction debris.
The formation of self-gravitating objects in tidal tails was already announced by \citet{zw56}.
However some interest in these objects has started growing only in recent years. Numerical 
simulations of galaxy encounters support two possible scenarios of tidal dwarf formation. 
The first 
one predicts the formation of massive bound condensations dominated by the stellar component 
along tidal tails and a subsequent infall of gas from the tail into the dwarf's potential well 
\citep{bh92,bh96}. The second one, based on numerical simulations including 
dissipation 
for the gaseous component, proposes that the gas condenses first \citep{eletal93}. Actually, both 
the stellar and gaseous component are observed in tidal tails and condensations along them.
The formation of condensations of gas and stars has been observed especially at the tips of tidal 
tails. A spectacular example is the one of the ``Antennae''  (NGC4038/4039, Schweizer, 1978; Mirabel, 
Dottori \&\ Lutz 1992), but other examples are shown e.g. in \citet{detal00} and \citet{wetal00}.
These ``condensations'' in tidal tails of interacting galaxy systems are commonly known as Tidal Dwarf 
Galaxies (TDGs). They usually share the typical properties of dwarf irregulars and blue compact 
dwarfs, concerning their surface brightness, blue colors, and strong bursts of star formation, 
but show somewhat higher metallicities confined in the narrow range 12+$\log$(O/H) $\sim$ 8.4 - 
8.6 \citep{detal00}. This is consistent with their formation from material already processed and 
chemically enriched in the outer disk of the progenitor galaxies and can be a clue to disentangle 
dwarf galaxies born as TDGs from the population of field dwarf galaxies. A more restrictive 
definition of TDGs has been proposed by \citet{detal00}, who consider a TDG ``an object which 
is a self-gravitating entity, formed out of the debris of a gravitational interaction''. This 
restricts such a category of objects to the ones having their own dynamics and a potential well 
deep enough to allow them to survive disruption for at least 1 Gyr.

Actually N-body simulations suggest that a substantial fraction of the material ejected during 
an interaction and accumulated in the tidal tails will slowly fall back into the remnant, 
allowing only for the outermost part (e.g., amounting to 20\% in the simulation of NGC7252 by 
\citet{hm95}) to gain enough kinetic energy to escape and evolve 
independently for a long time. Studies of poor galaxy groups have revealed
the presence of a population of dwarf galaxies in them \citep{zm98}, sometimes considerably 
increasing the number of group members. Also isolated elliptical galaxies that are considered to be 
relics of compact groups (CGs)
are surrounded by a population of dwarf galaxies \citep{mz99}. A possible explanation is 
that part of the dwarf galaxies observed in groups is generated in the groups themselves as a 
by-product of galaxy interactions in these dense environments. A search for TDGs in CGs led 
Hunsberger, Charlton, \& Zaritsky (1996) to suppose that up to half of the total dwarf 
population 
in CGs may be the product of interaction between galaxies. However the study of \citet{huetal96} 
suffers from the lack of kinematic information, which makes it difficult to distinguish 
``real'' TDGs from non-self-gravitating condensations in tidal debris. More restrictive 
identification criteria applied by \citet{ipetal00} led to the detection of a much lower number 
of TDGs in CGs.
However the study with most complete information on compact groups' TDGs until present is
the one by \citet{mdoetal01}, which relies on both photometry and kinematics, but is limited to 
HCG 92. They identified in HCG 92 seven candidate TDGs, at least two of which are thought to 
be able to survive fall-backs into or disruption by the parent galaxy.

In this paper we focus our attention on the brightness enhancements and/or blue 
condensations we identified as possible TDG candidates during a 
detailed analysis of the compact galaxy group CG J1720-67.8 
(Weinberger, Temporin, \& Kerber 1999 (Paper~I); Temporin et al. 2003 --hereafter Paper II).
In particular we perform optical and near-infrared aperture photometry 
and investigate the physical properties of the identified knots and candidate
TDGs through photoionization models calculated with the code Cloudy 90.04 \citep{fe96}.
Considerations about the tidal-tail kinematics are expressed on the bases of 
spectroscopic data obtained for discrete regions along the tail in addition to 
integral field spectra sampling its northern part.
The optical colors of the candidate TDGs are compared with the
grids of spectrophotometric evolutionary synthesis models by Weilbacher et al. (2000).

\section{PRIOR OBSERVATIONS AND NEW DATA}

\begin{figure}
\plottwo{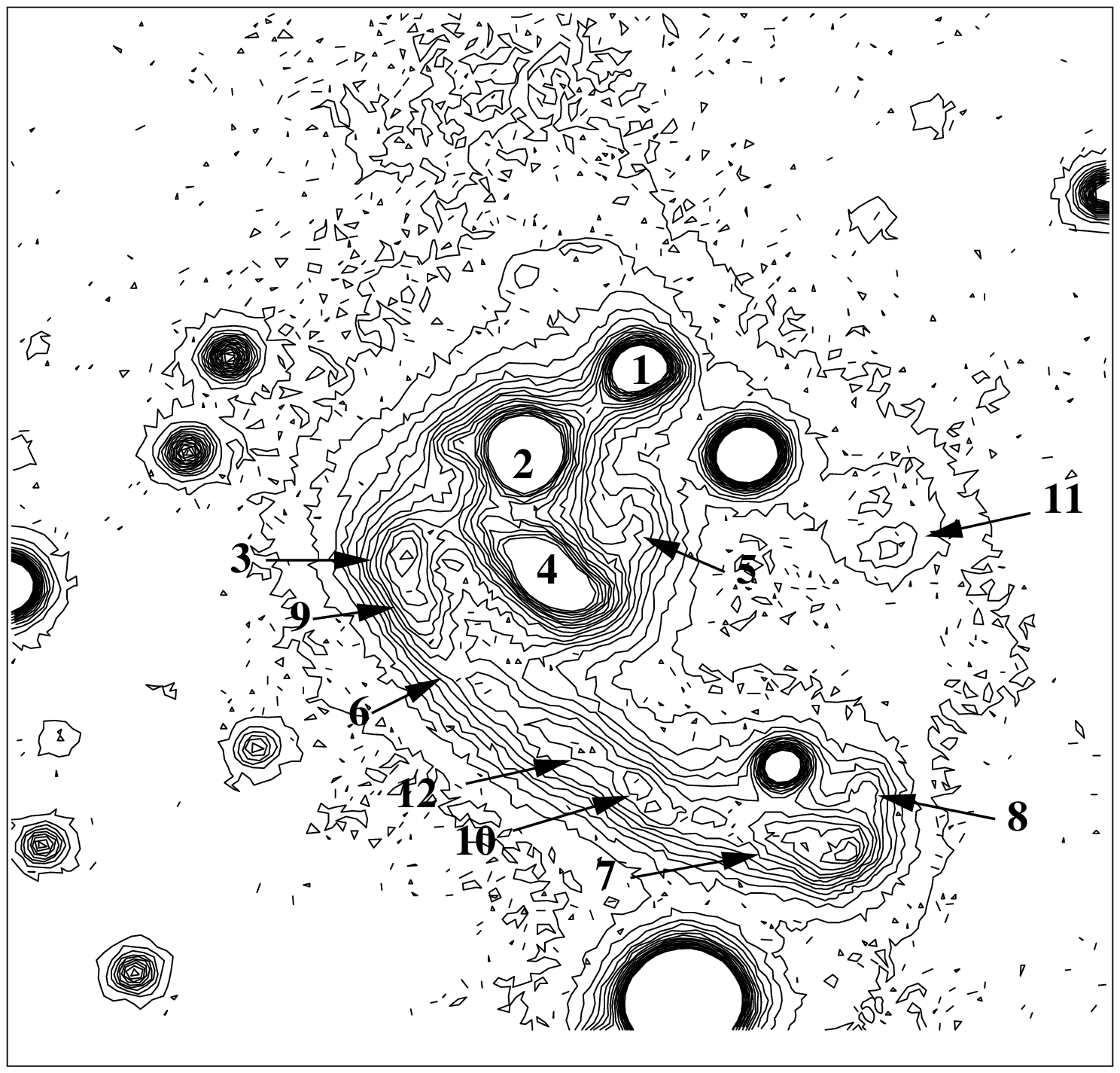}{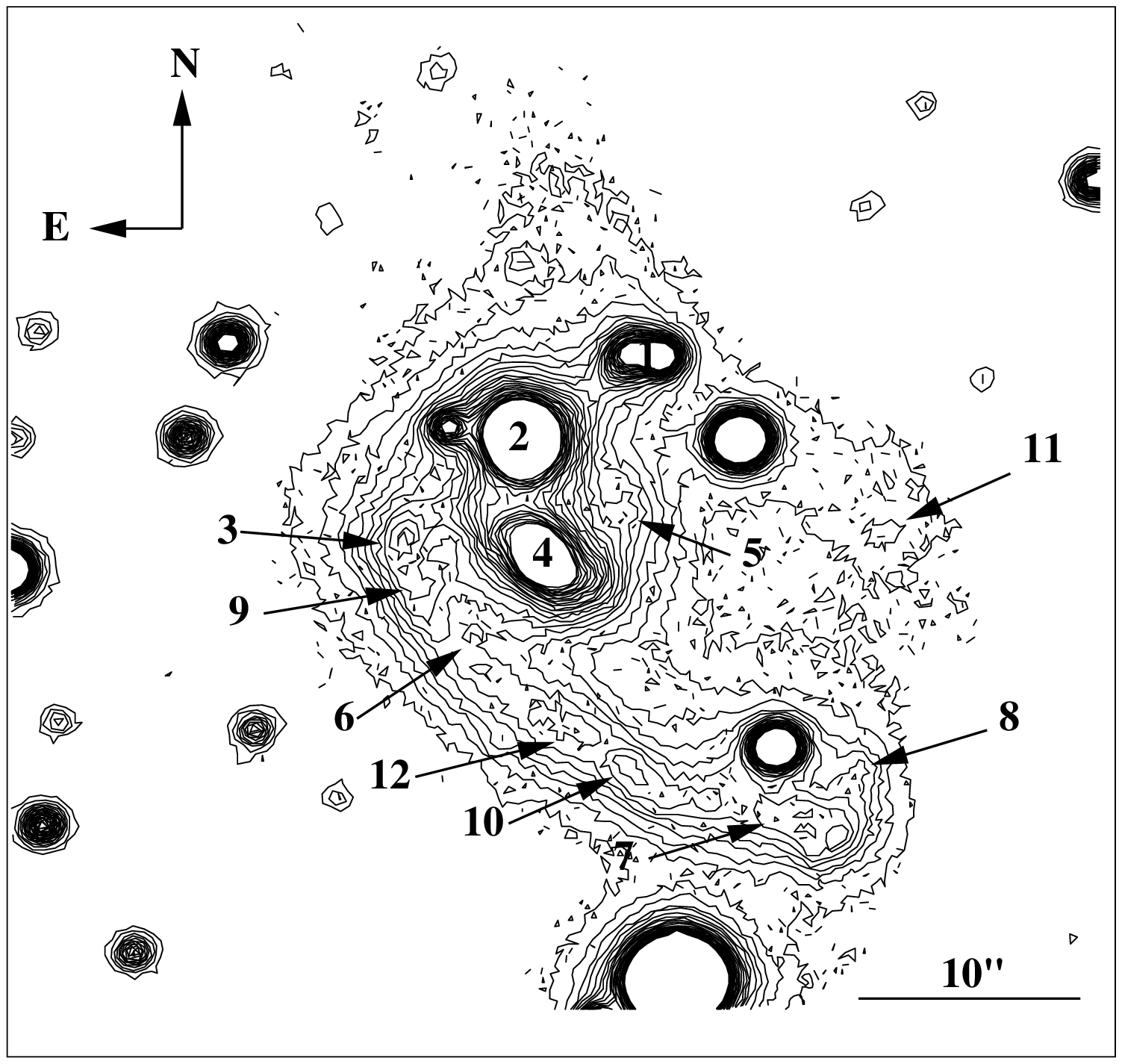}
\caption{Contour maps of the $B$ and $R$-band images of CG~J1720-67.8  
taken at the ESO 3.6 m telescope. The galaxies, TDG candidates and emission knots are 
labeled. Position no. 6 is a portion of the tidal arc that does not correspond to 
any visible knot and for which a spectrum is available (see Fig.~\ref{tdgspec}).\label{cmap}}
\end{figure}

Most of the data, on which the present study is based, have already been presented in Paper~II, 
therefore we only briefly list here the observational material at our 
disposal. Broad band $B$, $V$, and $R$ images (900 s, 600 s, and 3$\times$600 s exposures, 
respectively) have been obtained in June 1999 at the ESO 3.6 m telescope in La Silla. 
A number of broad band images was obtained in the near-infrared (NIR) regime in June 2000 at 
the 1 m Swope telescope in Las Campanas. The total integration times were 1800 s in the $J$ 
band, 2000 s in $H$ and 2500 s in $K$-short ($K_{s}$). Details about the reduction steps and the 
photometric calibration are given in Paper II. The NIR photometric system is described in 
detail in \citet{petal98}. 
For clarity we show in Fig.~\ref{cmap} the identification of the candidate TDGs and knots 
onto the contour maps of the $B$ and $R$-band images of CG J1720-67.8. 
Spectra of the candidate TDGs were obtained at the 2.5 m 
Du Pont telescope of Las Campanas with the Modular Spectrograph and at the ESO 3.6 m telescope 
in La Silla, with EFOSC2, in multi-object mode (MOS). They all show 
emission lines with radial velocities in agreement with those of the three main galaxies
and \ion{H}{2}-region like diagnostic ratios (see Table~6 and Fig.~10 in Paper II), although 
high values of the [\ion{S}{2}]/H$\alpha$ ratio suggest that shock heating might also be important.
Candidate TDGs' fluxes, the extinction
corrected intensities and the estimates of the internal extinction obtained from the Balmer decrement
were reported in Table~3 of Paper~II (but see also \S~4 below). 
The highest extinction value was found for object no. 3, the
candidate TDG at the northern tip of the group's tidal arc (E(\bv) = 0.78 $\pm$ 0.42).  

Additionally, we make use of velocity information about object 3+9 obtained from
recent observations (June 2002) at the 3.9 m Anglo-Australian Telescope with the integral field
unit SPIRAL (Segmented Pupil/Image Reformatting Array Lens) and a 600 g/mm grating in the 
wavelength range $\lambda$ 6500 - 7120 \AA\ in combination with the 2048$\times$4096 13.5 $\mu$-sized
pixels EEV chip. 

\placefigure{cmap}

\section{PHOTOMETRIC ANALYSIS}

From an examination of the contour maps in Fig.~\ref{cmap}, we note that objects 
3, 7, 10, and 12 along the tidal arc --selected for showing peaks of intensity
over the tidal feature where they are embedded-- show some sort of boundaries.
No. 9 appears like a secondary peak inside the boundaries of no. 3 --although
this could be an effect of insufficient spatial resolution-- therefore
we consider it as part of object 3 and we will refer to it as object 3+9. 
No. 8 appears like
a secondary peak adjacent to no. 7 and, although it does not show a clear 
brightness enhancement with respect to the tidal tail, it emerges as a sort of
separated knot in the $R$-band image after the application of an adaptive Laplacian
filter (Fig.~\ref{laplace}), which has the effect of suppressing the noise and 
enhancing faint structures\footnote{The position of faint structures revealed in close proximity of
bright sources appears slightly shifted in the filtered image with respect to the
original because of boundary effects and depending on the size of the filter adaptive box. 
Therefore object positions were defined only on the original images.} \citep{ri91,lo93}.
No. 5 emerges more clearly as a knot in Fig.~\ref{laplace}, while it appears as a 
diffuse structure connected to galaxy 4 in the contour maps of Fig.~\ref{cmap}.
This object could be a star-forming small tidal tail or perhaps a vestige of a
spiral arm. Its color does not differ from that of galaxy 4, as can be seen
in the \bv color map of Fig.~\ref{tdg_col} (obtained as described in Paper~II).
Finally, no. 11 appears as a knot
inside a ring-like (Fig.~\ref{laplace}) diffuse structure without any
obvious connection with tidal tails. 
In Fig.~\ref{tdg_col} all the objects identified along the tidal arc appear
bluer than the underlying tidal tail and the adjacent galaxy 4.

\placefigure{laplace}
\placefigure{tdg_col}

Since the features under investigation have irregular elongated shapes, we did
not apply traditional circular aperture photometry, which would include too much
flux from the surrounding parts of the tidal tail. Instead, we adopted a procedure
analogous to the one applied by \citet{wetal00} and measured total magnitudes
inside polygonal apertures. The apertures were
defined on the contour plot of the calibrated $B$-band image following the faintest
contour, which still allowed to separate each clump from the underlying tidal tail.
Therefore the surface brightness level at which the aperture was defined was different
for each object and the values are listed in Table~\ref{size}, together with peak
surface brightnesses. 
The same surface brightness levels were taken as reference to measure the size of
the clumps. Due to the elongated shape of the objects, instead of a typical radius
we list in Table~\ref{size} their minor and major dimensions in kpc, adopting
H$_0$ = 75 km s$^{-1}$ Mpc$^{-1}$.
The same polygonal apertures were applied to the aligned images in the 
$V$, $R$, $J$, $H$, and $K_{s}$ bands. The NIR images were first re-binned to match the 
pixel scale of the optical ones.
We chose not to subtract the tidal tail contribution from the objects' fluxes, but only
the sky-background evaluated in several positions as near as possible to the objects
of interest. This choice will allow us to compare our photometric data with the 
evolutionary synthesis models by \citet{wetal00}, which take into account also the old
stellar population present in the underlying tidal tail.

The total magnitudes and average surface brightnesses inside the polygonal apertures
and the optical and NIR colors
of the clumps are listed in Table~\ref{mag}. All the values are corrected for
Galactic extinction (E(\bv) = 0.088). 
We find optical colors (\br $\sim$ 0.65 - 0.85) consistent with those of HCG~92's TDG
candidates \citep{mdoetal01}.
Due to the difficulty to define a clear boundary for the knot no. 5 we give only
its surface brightness and colors, but not its total magnitude.
For comparison purposes we list also the average surface brightness and colors of
two portions of the tail measured between the clumps. The brightness of the tail 
increases slightly from North to South. 
Typical photometric errors in total optical magnitudes range from $\sim$
0.002 to 0.01 mag, therefore the calibration error ($\sim$ 0.05, see Paper~II) is dominating.
In the NIR photometric errors are typically $\sim$ 0.1 - 0.2 mag but reach 0.6 mag 
for the faintest objects, nos. 8 and 11, which could not be measured in the K$_s$ frame.

\placetable{size}
\placetable{mag}

Although for the candidate TDGs nos. 3+9 and 7 we could estimate a significant amount
of internal extinction from the Balmer decrement (see Paper~II, Table~3), the error
in this extinction measurement is high, therefore we preferred not to apply any 
internal extinction correction to the magnitudes of these objects. 
We obtained the following luminosities of the candidate
TDGs (including the light from the underlying tidal tail): 
M$_{B}$(3+9) = $-$17.46, M$_{B}$(7) = $-$17.06, M$_{B}$(8) = $-$13.93,  
M$_{B}$(10) = $-$15.86,
M$_{B}$(12) = $-$15.35, and M$_{B}$(11) = $-$14.39. 
We note that the above luminosities
are comparable or higher than those found by \citet{mdoetal01} in the candidate
TDGs' sample of Stephan's Quintet and those of TDG candidates in interacting systems
identified by \citet{wetal00}. 
The two brightest TDG candidates have luminosities higher than the brightest \ion{H}{2}
regions measured in Sc galaxies by \citet{bk97} and more typical of giant extragalactic \ion{H}{2}
regions \citep{ma94}. However, the comparison with \ion{H}{2} region properties 
is made difficult by the different methods of measurement used in the present work and 
in the literature. In particular, we need to emphasize that, unlike most published photometric
measurements of extragalactic \ion{H}{2} regions, we have used interactively defined polygonal
apertures and we did not subtract the flux contributed by the tidal tail, where the knots are
embedded, as explained above. We will further discuss this point later, in \S~5. 

\subsection{Other ``Knots'' in the Close\\ Environment of the Group}

A number of additional irregularly shaped, faint knots have been identified in
the close environment of CG~J1720-67.8, and in particular inside its halo.
Their nature is unclear and we have not yet obtained spectral information on them.
However, at least for part of them, we could measure aperture magnitudes in the optical images.
In this case we used circular apertures and evaluated the sky background in an
annulus around each object.
These knots are labeled with small-case letters onto the Laplacian filtered R-band
image in Fig.~\ref{laplace}.  
Their optical magnitudes and colors are listed in Table~\ref{knots}. 

\placetable{knots}

Some of these objects might belong to a population of dwarf galaxies associated with the
compact group, however deeper data and spectral information are necessary to understand it.
At present, the group's members among these cannot be distinguished from background objects
due to the lack of radial velocity information, although their colors could be consistent
with background galaxies (see \S~6).

\section{SPECTRAL PROPERTIES AND\\ PHOTOIONIZATION MODELS}

\placefigure{tdgspec}
The spectra of the candidate TDGs (nos. 3, 7, 8, 9, 10) and of additional knots of CG~J1720-67.8
(nos. 5, 11) all show a blue continuum (Fig.~\ref{tdgspec}), and some of them exhibit 
remarkable Balmer absorption lines in addition to the emission lines typical of \ion{H}{2} regions.
Therefore all their spectral properties agree to indicate a dominating young stellar population
and an activity of current/recent star formation. As already noticed in Paper~II, 
Veilleux \& Osterbrock's (1987) (VO-)diagnostic diagrams applied to these objects confirm that 
their gas is
photoionized by thermal sources --although a certain degree of ionization by shock-heating cannot
be excluded. Also the spectrum of the portion no. 6 of the arc shows Balmer absorption lines
and weak emission lines (H$\alpha$, [\ion{N}{2}], and [\ion{S}{2}]), indicating that a low level
of star formation is present in the tidal tail as well, together with an older stellar population.

Given the thermal nature of the ionizing source, the H$\alpha$ emission-line luminosity, after 
correction for internal extinction,
could be used to evaluate the star formation rate (SFR) of these objects.
In the spectra nos. 5 and 6, only a few emission lines were detected, i.e. the 
blend H$\alpha$-[\ion{N}{2}], the [\ion{S}{2}] doublet, and --for object 5-- [\ion{O}{2}] $\lambda$ 3727.
These were insufficient to locate the regions in the VO-diagnostic diagrams and the lack of
a measurable H$\beta$ prevented the estimate of the internal extinction. 
In this case we used the uncorrected H$\alpha$ luminosities for the SFR estimate.
The star formation rates calculated following \citet{hg86} are listed in Table~\ref{sfr}.
They range from 0.004 to 0.82 M$_{\odot}$ yr$^{-1}$,
with star formation rate densities (SFRD) of 0.05 to 14.15 M$_{\odot}$ yr$^{-1}$ pc$^{-2}$,
the highest value being found for the candidate TDG no. 3\footnote{The emission line fluxes of object no. 3, as well
as the extinction value, are the
ones reported in Paper~II, Table~3. These values are somewhat different from those reported in Paper~I, Table~1
because of a redefinition of the aperture in which they were measured. Additionally, we point out a mistake 
in Paper~I, Table~2, by which all the values referred to this object are too low.} 
and the lowest for knot no. 5.
These SFR values are comparable to those of the TDGs identified in HCG~92 \citep{mdoetal01}.

\placetable{sfr}

The observed emission-line ratios were used to infer the physical and chemical properties
of the ionized gas. 

TDGs are expected to have higher metallicities than typical field dwarf galaxies, 
since they form out of already enriched material ejected by the parent galaxies. 
Since in none of the candidate TDGs the [\ion{O}{3}] $\lambda$ 4363 emission-line was detected,
a direct measurement of the electronic temperature of the emitting gas --necessary for the 
calculation of the metal abundances of the gaseous component-- was not possible.
Therefore, in analogy with the method adopted for the three main galaxies
of CG~J1720-67.8 (Paper II), we obtained a first estimate of the 
metal abundances of the candidate TDGs by comparing their emission line ratios 
with the empirical diagrams of \citet{mg91} and \citet{dtt01}. 
From the comparison with the grid of models of \citet{mg91} in the plane [\ion{O}{3}]/[\ion{O}{2}]
\emph{vs} R$_{23}$ (R$_{23}$ = $\log$[([\ion{O}{2}] $\lambda$ 3727 + 
[\ion{O}{3}] $\lambda$ 4959,5007)/H${\beta}$]) we found that all the candidate TDGs have
an ionization parameter in the range $-$3.5 $\lesssim$ $\log$U $\lesssim$ $-$3. The 
metal abundances, represented by the O/H ratio, result in the range $-$3.9 
$\lesssim$ $\log$(O/H) $\lesssim$ $-$3.5 
for objects nos. 7, 8, and 9 and have the two possible values
$\log$(O/H) $\simeq \, -$4.1 or $\log$(O/H) $\simeq$ $-$3.2 for no. 10. 
Object no. 3 could not be compared with these model grids, because the [\ion{O}{2}] $\lambda$ 3727
was not detected in its spectrum.
All the candidate TDGs could be compared with the monotonic relation 12+$\log$(O/H) \emph{vs}
$\log$([\ion{N}{2}] $\lambda$ 6583/H$\alpha$) \citep{dtt01}. By referring to the model track
with $\log$U = $-$3.0 \citep[Fig.~1]{dtt01}, we found metallicity values in rough agreement with the 
lowest values given above --i.e. with the lower branch of the R$_{23}$ - $\log$(O/H) relation 
in \citet{mg91}-- namely $\log$(O/H) $\sim$ $-$3.7 for objects 3, 8, and 9, $\log$(O/H) $\sim$ $-$3.8 
for object 7, and $\log$(O/H) $\sim$ $-$3.9 for object 10, or, in other words, Z $\sim$ 0.2 - 0.3 
Z$_{\sun}$\footnote{As values for solar abundances we used O/H = 7.41$\times$10$^{-4}$, N/H = 
9.33$\times$10$^{-5}$, and S/H = 1.62$\times$10$^{-5}$.}.
These estimates were further refined by fitting the observed emission line ratios with 
photoionization models by use of the code Cloudy~90.04 \citep{fe96}.
During the fits the hydrogen density N$_{\rm H}$ and the ionization parameter U were let free to vary,
while as a typical temperature of the thermal ionizing continuum we assumed T = 4$\times$10$^4$ K.
Several models were tried with slightly different metallicity values, until the reproduction of
the observed line ratio was satisfactory (we chose the models with lowest $\chi^2$).
The final photoionization model parameters are given in Table~\ref{photoion}, while in 
Table~\ref{obs-mod} the observed and modeled emission-line intensities relative to H$\beta$ are 
compared. However, also with this procedure the metal abundances of objects 7, 8, and 9 could not
be uniquely determined. Actually photoionization models with physical parameters similar to those
given in Table~\ref{photoion} but
somewhat lower metal abundances would offer a comparably good, or even better fit to the observed
line ratios. The parameters of these alternative models are shown in Table~\ref{alt-photoion}.

\placetable{photoion}
\placetable{obs-mod}
\placetable{alt-photoion}

To summarize, our candidate TDGs show recent/present star formation activity, and their gaseous 
component exhibits an ionization parameter U in the range $\sim$ 4$\times$10$^{-4}$ - 
7$\times$10$^{-4}$, a hydrogen density N$_{\rm H}$ in the range $\sim$ 1 - 2$\times$10$^2$ cm$^{-3}$, 
and metal abundances 
Z $\sim$ 0.1 Z$_{\sun}$ for object 10 and Z $\sim$ 0.3 Z$_{\sun}$ for objects 3, 7, 8, and 9,
although somewhat lower abundances for the last three might be possible.
Especially object 7 could have Z $\sim$ 0.1 Z$_{\sun}$,
but with an overabundance of oxygen and sulphur, which appear to be roughly a third of the solar value.
As a comparison we recall here the metal abundances obtained for the three main galaxies 
of CG~J1720-67.8 (Paper~II): 
Z $\sim$ 0.2 Z$_{\sun}$ for galaxy 1, 
Z $\lesssim$ 0.5 Z$_{\sun}$ for galaxy 2, 
and Z $\sim$ 0.1 Z$_{\sun}$ for galaxy 4. 

\subsection{Velocity Field of the Northern Tip of the Tidal Arc}

During June 2002 observations at the Anglo-Australian Telescope equipped with SPIRAL we obtained
integral field spectra of the upper part of the tidal arc, where the candidate TDG 3+9 
is located. The object was covered with an array of 14$\times$15 microlenses giving
a field of view of 9$^{\prime\prime}$.8$\times$10$^{\prime\prime}$.5 with a spatial scale of 
0.7 arcsec per microlens per pixel. The position of the array onto the R-band image of the group
is shown in Fig.~\ref{array} (left). The observations were carried out with the nod and shuffle 
technique
in order to provide an optimal subtraction of the sky-background. The total exposure time was
1 hour, half of which was spent on target and the other half on sky.
The wavelength range was centered on the redshifted H$\alpha$ line. The 600R grating combined
with the EEV CCD-chip gave a dispersion of 0.612 \AA\ pixel$^{-1}$. The average spectral resolution 
was in the range
1.8 - 2.3 \AA\ with the lowest values in the central part of the array and highest values at the borders.
The basic reduction steps --involving the tracing of the fibres by use of a flat-field frame, bias
subtraction, flat-fielding, wavelength calibration, sky subtraction, and combination of subsequent
exposures-- were done with the dedicated software 
2dfdr\footnote{Available at http://www.aao.gov.au/2df/software.html}, Version 2.3, provided 
by the Anglo-Australian Observatory.
This software allows also the image reconstruction at a specified wavelength or in a specified 
wavelength range and was used to reconstruct an H$\alpha +$continuum image of the field in the range
6800 -- 6900 \AA\ for object identification purposes. 
This image was resampled onto a grid of
56$\times$60 pixels (i.e. 4 times larger than the original) and smoothed with a 4$\times$4 pixel 
box for viewing purposes (Fig.~\ref{array}, right).

Further analysis of the data was carried out within IRAF. Specifically, for each spectrum where 
the H$\alpha$ emission line was detected we obtained a radial velocity measurement (through Gaussian
fit of the emission-line) in order to
reconstruct the velocity field of the source. 
The average calibration error, evaluated from position measurements of OH night-sky lines 
\citep{ost96}, is 2.9 km s$^{-1}$. No systematic trends of the error were detected across 
the 14$\times$15 array of spectra.
The result is shown in Fig.~\ref{vfield}. On the
left-hand panel the original reconstructed continuum image (without resampling) shows object 3+9 in the
center and portions of galaxies 2 and 4 at the right edge. On the central image the velocity field,
with indicated the maximum and minimum measured radial velocities of the object 3+9, shows 
a velocity gradient in direction NE-SW (the north-eastern part being the receding one) across the
object, with a maximum velocity difference of $\sim$ 200 km s$^{-1}$ over an extent of $\sim$ 5 kpc.
The velocity field is shown also after magnification and resampling (Fig.~\ref{vfield}, right-hand panel) 
and can be compared with Fig.~\ref{array}.
There might be several possible explanations for such a gradient \citep[see e.g.][]{weil02}, like
streaming motion along the tidal tail, projection effects, or even rotational motion.
We lack detailed information on the radial velocity gradient along the tidal tail, apart from
the discrete measurements in correspondence of objects 7, 8, and 10 that indicate that the southern 
part of the tail is approaching us (radial velocity $\sim$ 13200 km s$^{-1}$) and showing a velocity 
difference $\lesssim$ 400 km s$^{-1}$ between the tips of the tail.
However, the TDG candidate 3+9 is located at the base of the tail, where tidal streaming and geometric 
effects are most extreme \citep[see e.g. the case of the merger remnant NGC 7252][their Fig.~1]{hm95}.
Such effects are likely to be responsible for the observed gradient.
Even though, the presence of a rotational motion cannot be ruled out. 
If the velocity gradient of object 3+9 was actually entirely caused by rotation, it would imply a dynamical 
mass M$_{\rm dyn} \, \sim$ 6$\times$10$^9$ M$_{\odot}$, comparable to the mass estimated
for the group's galaxy no. 1 \citep{tfva01}. This mass, combined with the $B$-band luminosity
given in \S~3, would yield a mass-to-light ratio M$_{\rm dyn}$/L$_{B}$ $\sim$ 4.
However, also in case of self-gravitation, the object is 
unlikely to be in a relaxed state, therefore the application of the virial theorem could lead to 
a considerable overestimate of its mass.

\placefigure{array}
\placefigure{vfield}

\section{COMPARISON WITH PROPERTIES\\ OF EXTRAGALACTIC\\ \ion{H}{2} REGIONS}

In the attempt to understand the real nature of the knots in CG~J1720-67.8 and to investigate
alternative possibilities to the TDG candidate hypothesis, we compare their optical properties
with those of normal and giant extragalactic \ion{H}{2} regions (GEHR) present in the literature.
As already stated in \S~3, some difficulties in such a comparison arise as a consequence of
different methods of measurement.
In particular the background subtraction is critical when comparing photometric properties.
Since most of the published optical photometry of \ion{H}{2} regions is obtained by subtracting
the contribution of the underlying galaxy components (i.e. disk, spiral arms), we have 
estimated the contribution of the tidal tail to the light of our TDG candidates in the $BVR$ bands
and subtracted it from the measured fluxes of the knots.
The new set of background-subtracted magnitudes and colors is shown in Table~\ref{no_bkg}.
The background of each knot was estimated by averaging the mean value inside a number of 10$\times$10
pixel boxes selected all around the polygonal aperture used for the photometry, taking care to avoid
the inclusion of adjacent knots.
We stress that the background around the knots has strong gradients, especially
at the tips of the tidal tail. Therefore, although an
as large as possible number of boxes was used for its estimate, the final values we adopted have 
uncertainties in the range $\sim$ 12 -- 40 per cent. 
We found that the tidal tail contributes to the flux in the clumps by $\sim$ 45 to 60 per cent in $B$,
55 to 68 per cent in $V$, and 52 to 70 per cent in $R$. The highest percentages are found in the 
central parts of the arc, while the lowest at the tips of the arc. The maximum contribution from the
tail is observed in the $V$ band and the minimum in the $B$ band. Such a differential contribution
determines the difference between the colors given in Table~\ref{mag} and those in Table~\ref{no_bkg}.

The \bv and \vr\ colors given in Table~\ref{no_bkg} can be compared with those measured by \citet{ma94}
 for GEHR in a sample of nearby spiral and irregular galaxies.
 Median colors of GEHRs and GEHR groups in Mayya's sample are \bv = 0.21, \vr = 0.48 and \bv = 0.30,
 \vr = 0.44, respectively, which are consistent with the values we found for our clumps.
Actually in \citet[][Table~3]{ma94} there are several examples of GEHRs with colors very similar to
those in our Table~\ref{no_bkg}. However, their H$\alpha$ equivalent widths (W$_{\rm H\alpha}$) are 
much higher than the ones we measured in the spectra of our clumps (Table~\ref{sfr}). 
Furthermore the sizes of GEHRs 
are $\lesssim$ 1.5 kpc, while the clumps in CG~J1720-67.8 are considerably more extended (see 
Table~\ref{size}).

The properties of the first ranked \ion{H}{2} regions in a sample of nearby spiral and irregular
galaxies were studied by \citet{k88} through H$\alpha$ photometry.
The brightest (``giant'' or ``supergiant'') \ion{H}{2} regions in Kennicutt's sample were found almost
exclusively in late-type normal galaxies or in peculiar galaxies, and the question ``whether they simply
represent the high-luminosity tail of the normal \ion{H}{2} region luminosity function or are a physically
distinct class of objects formed under special conditions'' appears to remain unanswered.
H$\alpha$ luminosities of these regions are in the range 10$^{39}$--10$^{41}$ ergs s$^{-1}$ and the masses
of the embedded star clusters, calculated assuming a Salpeter IMF and including only stars with masses
10--100 M$_{\odot}$, were expressed as M$^{\bigstar}$ = 3600 L$_{39}$ M$_{\odot}$ 
(where L$_{39}$ is the H$\alpha$ luminosity in units of 10$^{39}$ ergs s$^{-1}$), while 
the ionized gas mass was found to scale roughly linearly with H$\alpha$ luminosity (at least for the 
largest \ion{H}{2} regions), although the actual value depends on the density $N_e$ according to
the relation M$_{gas}$ = 2.3$\times$10$^6$ L$_{39}$/$N_e$.

By applying the above relations to our TDG candidates --using their extinction-corrected, 
spectroscopically determined, H$\alpha$ luminosities given in Table~\ref{sfr} and the hydrogen densities 
given in Table~\ref{photoion}-- we find masses of the embedded clusters ranging from $\sim$
4$\times$10$^4$ M$_{\odot}$ (object 10) to $\sim$ 5$\times$10$^5$ M$_{\odot}$ in the brightest knots at 
the tips of the tail and ionized gas masses ranging from $\sim~10^5$ (object 10) to $\sim$ 
4$\times10^6$ M$_{\odot}$ (object 3+9). 
These are only rough estimates and the actual mass values could be larger, since we have used here
H$\alpha$ luminosities from long-slit spectra, which covered big fractions, but not the full extent of
the objects. For comparison, the H$\alpha$ luminosity of the position 6 along the tidal tail would
yield M$^{\bigstar}$ $\sim$ 10$^4$ M$_{\odot}$ and M$_{gas}$ of order 10$^5$ M$_{\odot}$ (in this 
case the density could not be determined and the ionized gas mass is estimated through \citet{k88}
empirical relation).
The equivalent ionizing luminosity of object 3+9 reaches 10$^{53}$ photons s$^{-1}$
(after correction for internal extinction). 

The properties of the knots along (and especially
at the tips of) the tidal arc are comparable to those of the giant and supergiant
extragalactic \ion{H}{2} regions of Kennicutt's sample. We note that H$\alpha$ luminosities given in
\citet{k88} are not corrected for internal extinction, unlike those in our Table~\ref{sfr}. 
However, also before the extinction correction our TDG candidates have $\log$ H$\alpha \gtrsim 40$ 
ergs s$^{-1}$. 

A comparison of H$\alpha$ equivalent widths given in \citet[][their Fig.~6]{k89} with 
those in Table~\ref{sfr} shows that our TDG candidates are located at the lower limit of the range
found for disk \ion{H}{2} regions. Furthermore the value measured in the spectrum no. 9 is extremely low
and similar to the values measured in spectra nos. 5 and 6. This could indicate a non-negligible
contribution to the stellar continuum from an underlying evolved stellar population.

\section{COMPARISON WITH TDG\\ EVOLUTIONARY SYNTHESIS\\ MODELS}

The lack of detailed kinematic information (except for the low-resolution velocity field 
of the ionized gas in object 3+9)
prevent us from establishing which, if any, of the candidate TDGs we have identified
are actually self-gravitating objects --i.e. TDGs in the most rigorous meaning of the term-- 
or simply clumps of gas and stars, which are still bound to the tidal tail where they have formed
and might eventually evolve into TDGs.
Even though, evolutionary synthesis models have proved useful to discriminate
between candidate TDGs or TDG progenitors and background objects, and also to estimate
the ratio of old to young stellar mass in them \citep{wetal00}.
Therefore we compare here the observational properties of our candidate TDGs with the
grids of evolutionary synthesis models published in \citet{wetal00}.
They have obtained models for two different metallicities, in the range expected for TDGs,
i.e. Z$_1$ $\sim$ Z$_{\sun}$/18 and Z$_3$ $\sim$ Z$_{\sun}$/2.3, by use of an evolutionary code
based on the work of \citet{kr95}. Their models assume for TDGs the same undisturbed evolutionary
history of the parent galaxies until the onset of the interaction-induced starburst.
A \citet{sc86} initial mass function is assumed. In the burst the SFR is set to a maximum value
and decreases exponentially with a timescale $\tau_{B}$.

The grids of evolutionary models obtained by \citet{wetal00} for different burst strengths are 
represented by counter-clockwise loops in the \bv $vs$ \vr\  two-color diagram.
Since we have found that our candidate TDGs have metallicities in the range
0.1 Z$_{\sun}$ $\lesssim$ Z $\lesssim$ 0.3 Z$_{\sun}$, and in particular oxygen abundances
O/H $\approx$ 0.35 (O/H)$_{\sun}$, we have compared their optical colors with the Z$_3$ models
from \citet{wetal00}. 
The observed colors of our candidate TDGs are marked in Fig.~\ref{grid} onto the two-color
diagram adapted from Fig.~1c of \citet{wetal00}. Values corrected only for foreground
Galactic extinction are marked with crosses. 
Artificial data-points (labeled ``a'', ``b'', ``c'', ``d'', and ``e'')
from \citet{wetal00} are plotted, as well (full circles). In particular point ``b''
corresponds to a burst strength $b \, >$ 0.1 and a burst age $\sim$ 20 Myr,
point ``c'' indicates a medium to strong burst of much higher age ($\sim$ 80 Myr), while
point ``d'' corresponds to a burst age of $\sim$ 7 Myr or less, depending on the metallicity.
Most of our data-points fall into a region of the two-color diagram where tracks of different
models overlap, namely the part of the tracks with burst ages of only a few Myrs. 
This would indicate that such objects have just started
their bursts of star formation. However, the application of a correction for internal 
extinction would move the points in the diagram toward older burst ages, reaching the region
occupied by the artificial data-point ``b'', or it would imply stronger bursts.
Older burst ages would be in agreement with the observed H$\alpha$ equivalent widths (Table~\ref{sfr}),
which are not as high as expected and observed in case of very young (age $<$ 2 - 3 Myr) \ion{H}{2} 
regions \citep[see e.g.][]{bk97}. However, the Balmer emission-line equivalent width could be
lowered as an effect of the presence of an underlying population of stars
older than those responsible for the ionization and contributing to the stellar continuum.
Internal extinction affecting the emission lines but not the continuum could also play a role
in lowering the equivalent width \citep[e.g.][]{mp96}.

Although it is not possible to establish the appropriate
duration and strength of the bursts due to the similarity of models
with $b$ in the range 0.05 -- 0.18 and $\tau_B$ $\sim$
5$\times$10$^5$ -- 1$\times$10$^6$ yr,
object 8 appears consistent with a weaker burst with respect to the other TDG candidates.
Object 11, a condensation in a ring-like
structure on the west side of the group, falls near the artificial datapoint ``c'', 
suggesting a much higher burst age. This could be a condensation formed in an already 
fading tidal tail.

Supposing that these TDG candidates will evolve following the tracks in Fig.~\ref{grid}, they 
will become still bluer during the next few Myr, then they will become progressively redder,
reaching after $\sim$ 1 Gyr the pre-burst colors. Unless new bursts of star formation
are triggered, the objects will progressively fade during their evolution. Their fading can
be quantified by comparison with the luminosity evolution modeled by \citep{wetal00}, which 
shows a rapid luminosity decrease in the first 50 Myrs after the maximum of the burst. 
In the hypothesis that they have reached the maximum $B$ luminosity at the present epoch, they will
fade by about 3.5 mag within 2 Gyr. Therefore the brightest clump, no. 3+9, would reach 
M$_{B}$ $\sim$ $-$14 and it would be still observable, with an apparent magnitude $B$ $\sim$ 22.3.
In the same time, the faintest object, no. 8, would fade to $B$ $\sim$ 25.9 mag and would be visible 
only in very deep exposures.
Object, no. 11, according to its estimated burst age of $\sim$ 80 Myr, should have
already faded by $\sim$ 2.2 mag with respect to the luminosity peak reached during the burst phase
and its $B$ luminosity is expected to decrease by another 1.3 mag in the next 1.9 Gyr, down to 
$B$ $\sim$ 23.2 mag.

\placefigure{grid}

The observed colors of the additional knots `a', `b', `c', `d', and `e' (Fig. \ref{laplace}) 
identified in the group's halo
would fall far away from the TDG model tracks, therefore, according to \citet{wetal00},
they could be regarded as background objects.

\section{DISCUSSION AND CONCLUSION}

We have presented photometric and spectroscopic observational data of a number of knots and
candidate TDGs previously identified (Paper~I and Paper~II) at the ends of and along the tidal
features and in the halo of the ultracompact group CG~J1720-67.8. 
Although we are not able to 
establish whether the clumps along the tidal tail are already self-gravitating objects because of the lack 
of kinematic information --with the possible exception of object 3+9 whose velocity 
field shows a regular velocity gradient--  we have found that their properties are consistent with 
those expected for TDGs or TDG progenitors. Namely, they show blue colors, recent
and/or present-day star formation activity, and relatively high oxygen abundances
with respect to field dwarf galaxies.
The same properties would also be consistent with star formation within the tail. Indeed we
have detected a low level of star formation in a region of the tail where no visible knots
are present (position no. 6) and the moderately blue colors of the tail suggest that some
degree of star formation is present all along it. However the star-forming clumps we have identified
as candidate TDGs clearly appear as substructures and condensations within the tail and 
have colors significantly bluer than the surrounding regions. Moreover their luminosities are
considerably higher than those typical of normal \ion{H}{2} regions in spiral galaxies and
more typical of GEHRs. Optical colors as well, once a correction for contamination from 
the tidal tail material has been applied, are in agreement with those of GEHRs \citep{ma94}.
Actually, optical colors are not sufficient to discriminate between (giant) \ion{H}{2} regions
and TDGs.

The sizes of most of our candidate TDGs are considerably larger than those of the most extended GEHRs, 
which reach at most $\sim$ 1.5 kpc \citep[e.g.][]{k88}.
Instead they have dimensions typical of dwarf galaxies \citep[see e.g.][]{elm96}
and comparable to those 
of the candidate TDGs in Stephan's Quintet \citep{mdoetal01}.
Only object no. 8, judging from its size, might be interpreted as a giant \ion{H}{2} region.
At the seeing of our observations, the spatial resolution element is $\lesssim$ 1 kpc,
which means that a giant \ion{H}{2} region would barely be resolved, while the 
clumps we observe appear clearly extended. 
Some of them --e.g. nos. 10 and 12 that appear less concentrated than the objects at the tips
of the tail-- could be complexes of \ion{H}{2} regions, which 
appear as an individual clump due to insufficient spatial resolution.
However, if such complexes form in the tail, their potential well might result deep 
enough for them to become bound objects that  eventually can evolve as dwarf galaxies, 
provided their mass is sufficient. In such a case they could be considered as TDGs in the process 
of formation.
The presence itself of structures in the tail suggests that self-gravitation might play some role.
Whether or not the clumps of stars and ionized gas are to become bound objects and even accrete more
matter from their surroundings, depends on their having enough mass to survive fall-back
into the parent galaxies and tidal disruption \citep{hib94}.
 
The measured optical colors of our candidate TDGs are consistent with
evolutionary synthesis tracks calculated for TDGs in the expected metallicity range
\citep{wetal00}. The positions in the two-color diagrams, when compared with evolutionary tracks,
suggests that the objects
identified along the group's tidal arc have burst ages of a few Myr, while the blue
knot no. 11 embedded in a ring-like structure in the group's halo is consistent with a
much older burst age ($\sim$ 80 Myr). However, the location of the objects in the two-color
diagram might be strongly affected by internal extinction, which has not been taken
into account due to the high uncertainty in its determination. The effect could be relevant
especially for the objects at the tips of the tail, where indications of higher
extinction are found. The application of an internal extinction correction would move
the observational points toward higher burst ages (up to $\sim$ 20 Myr) in the model tracks,
thus offering a possible explanation to the relatively low observed H$\alpha$ equivalent widths.

If a luminosity evolution analogous to the one modeled by \citet{wetal00} is assumed, the
identified TDG candidates should fade to values of the $B$ luminosity still detectable after
2 Gyr of evolution, provided they survive as bound objects. 
 
A comparison of the metallicities of the candidate TDGs with those of the three main galaxies 
in the group does not easily allow to understand which one is the parent galaxy.
Actually object no. 10 is the one with lowest metallicity, in agreement with the metal 
abundances of galaxy 4 (Paper~II). For the other objects the metallicity value is
uncertain, although it seems established that their oxygen abundance is approximately
a third of the solar one, therefore higher than that of galaxy 4.
Their parent galaxy might be no. 1, which exhibits somewhat higher abundances.

A quantitative estimate of the possibility for these objects to survive as individual 
galaxies against internal motion and tidal forces exerted by the parent galaxies would
require an estimate of their virial and tidal masses \citep{mdoetal01}. The first cannot
be derived with the presently available data, the latter would require knowledge of the mass 
interior to their orbit. Since this mass is not known,
we limit ourselves to a few qualitative considerations.
We have found that the object composed of no. 3 and no. 9 (considered together) is the 
brightest of the proposed candidate TDGs. It's velocity gradient, if actually due to rotation,
would imply a mass comparable to the one estimated for galaxy 1 \citep{tfva01}. 
If this was the case, its mass could be sufficient for it to escape tidal 
disruption, despite its unfavorable position in the tidal feature,
very nearby to the parent galaxies. However, the observed velocity gradient could
be an effect of streaming motion in combination with projection effects, which can be
extreme at the base of the tidal tail. Therefore, no conclusive argument can be given at 
present.
Object no. 7, although less bright and therefore presumably less massive, seems to be 
in a more favorable location to survive
tidal disruption. 

In principle NIR photometry, when combined with optical photometry and equivalent width
measurements, should help at least to understand whether the modest luminosity 
enhancements of our TDG candidates with respect to the surrounding regions of the tidal tail 
indicate an
actual increase of the underlying mass density --as expected in case of distinct dynamical entities-- 
and not only an increase of the luminosity as a consequence of the burst of star formation. 
However the question cannot be addressed in 
a straightforward manner. Not only the optical (especially $B$) luminosity is influenced by the
recent star formation activity with consequent changes in the M/L ratio, but also the NIR luminosity
is affected, although in smaller measure. Therefore accurate evolutionary synthesis models, with
adequate time resolution, are necessary for a meaningful interpretation of the data. 
Additionally, the uncertainty in the burst strength and age determination,
partly caused by uncertainty in the internal extinction determination as explained above, further
complicates the matter. We will attempt to address this question in a future work.

At present, on the base of the analysis presented here, we can only conclude that the two objects
at the tips of the tidal tail (3+9 and 7) appear to be the most promising TDG candidates of the
galaxy group.

\acknowledgments
We thank the anonymous referee for useful comments and suggestions, which helped to considerably
improve this paper.
ST is grateful to P. M. Weilbacher for fruitful discussions.
RW is grateful to the Austrian ``Bundesministerium f\"ur Wissenschaft und Verkehr"
for travel support. GG thanks ``proyecto DIPUC 2001/14-E''.
ST acknowledges support by the Austrian Science Fund (FWF) under project no. P15065.

\clearpage
\onecolumn

\begin{figure}
\figcaption{Adaptive laplacian filtered R-band image of CG~J1720-67.8. 
Polygonal apertures roughly approximating those defined on the $B$-band image and used for aperture 
photometry (see text) are drawn around the knots and candidate TDGs already labeled in Fig.~\ref{cmap}. 
Note that objects 5 and 8 are clearly visible in the filtered image, but their position is shifted with 
respect to the apertures, which were defined in the original image. This is an effect of the 
filtering algorithm when faint structures are revealed in proximity of much brighter sources.
For photometric measurements only the original (non-filtered) image was used. 
The knots identified in the close environment of CG~J1720-67.8 and listed in Table~\ref{knots} 
are indicated with circles or ellipses 
and marked with small-case letters. Stars close to the group's members are marked with ``S''. 
\label{laplace}}
\end{figure}

\begin{figure}
\figcaption{\bv color map obtained after alignment of the $B$ and $V$ frames and matching of
their point spread functions, as described in Paper~II. All objects of the group are labeled
and the polygonal apertures used for the photometry of knots and TDG candidates are drawn. 
The color scale is selected in a way such
that physically bluer structures appear blue in the image and redder structures appear
red. \label{tdg_col}}
\end{figure}

\clearpage

\begin{figure}
\epsscale{.7}
\plotone{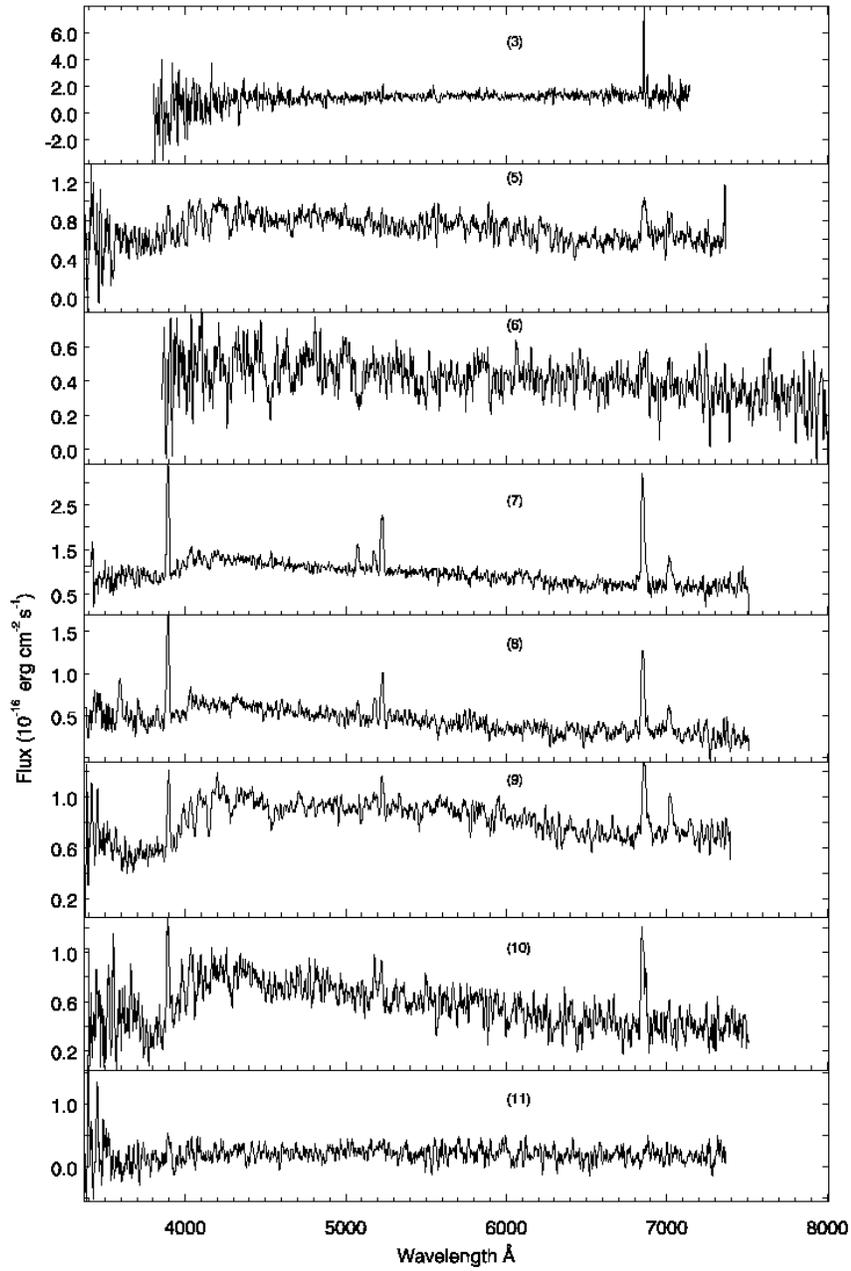}
\caption{Spectra of the knots and tidal dwarf candidates of CG J1720-67.8 \label{tdgspec}}
\end{figure}

\begin{figure}
\figcaption{Left: SPIRAL microlenses array position marked over the $R$-band image of CG J1720-67.8.
         Right: H$\alpha +$continuum map reconstructed from the array of SPIRAL spectra in the
	 range 6800 -- 6900 \AA. For viewing purposes the original map has been magnified, 
	 projected onto a grid of 56$\times$60 pixels, and smoothed with a 4$\times$4 pixel box.\label{array}}
\end{figure}

\clearpage

\begin{figure}
\figcaption{Left: H$\alpha +$continuum map reconstructed from the array of SPIRAL spectra in the
	 range 6800 -- 6900 \AA\ (the same as Fig.~\ref{array} but with the original pixel size).
	 Center: H$\alpha$ velocity field. Velocities increase from blue to red. The minimum and
	 maximum radial velocities measured on object 3+9 are indicated.
	 Right: The same as the central frame, after projection onto a grid of 56$\times$60 pixels.
	 North is on top, East to the left. \label{vfield}}
\end{figure}

\begin{figure}
\plotone{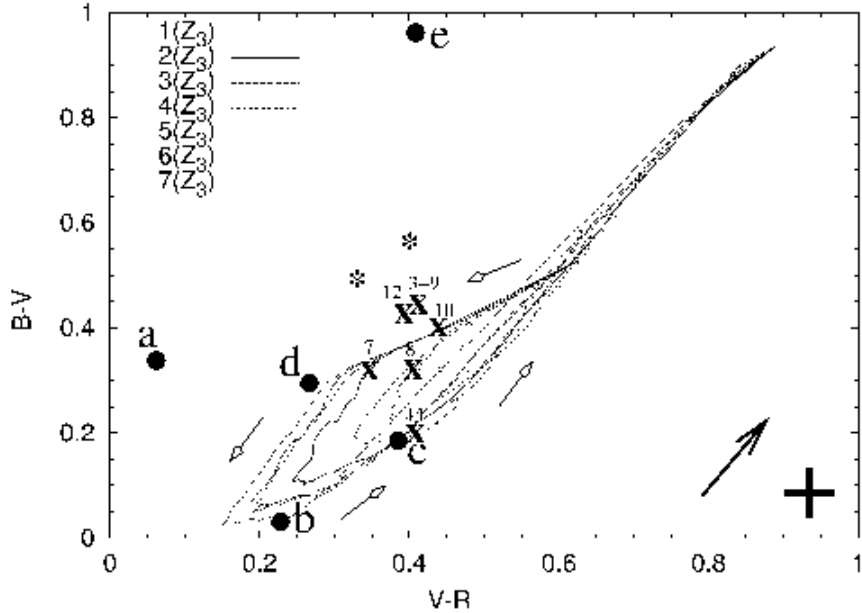}
\caption{Evolutionary synthesis grid of models for TDGs in the
\bv \emph{vs} \vr\ plane,
adapted from Fig. 1c of \citet{wetal00}, with overplotted 
data-points relative to the candidate TDGs of CG~J1720-67.8, marked with
`X'. For comparison, the positions of two regions of the tidal tail
measured between the knots are marked, as well ($\ast$). 
The cross in the lower right
corner shows the estimated typical error in colors, while the big arrow 
represents a reddening A$_{B}$ = 0.5 mag.
The smaller arrows around the color loops 
indicate increasing ages after the burst onset. Filled
circles are artificial data-points from \citet{wetal00}
modeling. The one labeled `b' corresponds to a burst strength $b$ $>$
0.1 and a burst age $\sim$ 20 Myr, `c' indicates a medium to strong burst
($b$ $>$ 0.05) and a burst age $>$ 80 Myr, while `d' corresponds to an age
$\sim$ 7 Myr. The artificial point `e' is an example of background galaxy.
Color loops for seven different burst strengths are shown. \label{grid}}
\end{figure}

\clearpage

\begin{deluxetable}{cccc}
\tabletypesize{\footnotesize}
\tablecaption{\label{size}Aperture and Peak Surface Brightness Levels and Clump Sizes}
\tablewidth{0pt}
\tablehead{
\colhead{Object ID} & \colhead{Aperture $\mu_{B}$}& \colhead{Peak $\mu_{B}$}&
\colhead{$a \times b$}}
\startdata
3$+$9& 22.35& 21.80 & 6.5$\times$2.0 \\
5  & 22.47 & \nodata & 2.0$\times$1.1\\
7  & 22.24 & 21.780 & 4.9$\times$2.5\\
8  & 22.48 & 22.35 & 1.3$\times$0.6\\
10 & 22.25 & 22.03 & 3.5$\times$1.4\\
11 & 23.54 & \nodata & 2.2$\times$1.7\\
12 & 22.26 & 22.14& 4.5$\times$1.7\\
\enddata
\tablecomments{Aperture and peak $\mu_{\rm B}$ are in mag arcsec$^{-2}$, 
$a$ and $b$ are the major and minor dimension of the clumps expressed in kpc; 
all surface brightness values are corrected for Galactic extinction.} 
\end{deluxetable}

\begin{deluxetable}{ccccccccccccc}
\tabletypesize{\footnotesize}
\tablecaption{\label{mag}Magnitudes, Colors and Average Surface Brightnesses}
\tablewidth{0pt}
\tablehead{
\colhead{Object ID} & \colhead{$B$ }& \colhead{\bv}& \colhead{\vr}& 
\colhead{\vh}& \colhead{\jh}& \colhead{\hk}&
\colhead{$\mu_{B}$}& \colhead{$\mu_{V}$}& \colhead{$\mu_{R}$} &
\colhead{$\mu_{J}$}& \colhead{$\mu_{H}$}& \colhead{$\mu_{K_s}$}}
\startdata
3$+$9& 18.72 & 0.44 & 0.41 & 2.16 & 0.66 & 0.25 & 22.07 & 21.63 & 21.22 & 20.13& 19.47& 19.22 \\
5  & \nodata & 0.58 & 0.43 & 2.22 & 0.74 & 0.21 & 22.40 & 21.81 & 21.38 & 20.33& 19.59& 19.38 \\
7  &   19.22 & 0.31 & 0.35 & 1.87 & 0.68 & 0.05 & 22.04 & 21.74 & 21.39 & 20.55& 19.86& 19.81 \\
8  &   22.35 & 0.31 & 0.40 & 2.12 & 0.88 & \nodata   & 22.44 & 22.14 & 21.73 & 20.90& 20.02& \nodata    \\
10 &   20.42 & 0.41 & 0.43 & 2.01 & 0.73 & 0.25 & 22.17 & 21.76 & 21.32 & 20.49& 19.75& 19.50 \\
11 &   21.89 & 0.19 & 0.41 & 1.87 & 0.90 & \nodata   & 23.36 & 23.17 & 22.75 & 22.20& 21.30& \nodata    \\
12 &   20.93 & 0.43 & 0.39 & 1.99 & 0.55 & 0.54 & 22.22 & 21.79 & 21.40 & 20.41& 19.80& 19.26 \\
arc$_1$ & \nodata & 0.56 & 0.40 & 1.74 & 0.73 & 0.29 & 22.47 & 21.91 & 21.51 & 20.42& 19.69& 19.40 \\
arc$_2$ & \nodata & 0.50 & 0.33 & 1.28 & 0.62 & 0.55 & 22.34 & 21.81 & 21.51 & 20.67& 20.05& 19.50 \\
\enddata
\tablecomments{All values are corrected for Galactic extinction. The measurements of two positions 
along the arc between the clumps are indicated with arc$_1$ and arc$_2$ from North to South.} 
\end{deluxetable}

\begin{deluxetable}{cccccc}
\tabletypesize{\footnotesize}
\tablecaption{\label{knots}Additional Knots: Optical Magnitudes and Colors }
\tablewidth{0pt}
\tablehead{
\colhead{Object ID} &  \colhead{$B$} & \colhead{$V$} &
 \colhead{$R$} & \colhead{\bv}& \colhead{\vr}}
\startdata
a & 24.33 & 22.03 & 21.22 & 2.30 & 0.82 \\
b & 20.06 & 19.32 & 18.85 & 0.74 & 0.46 \\
c & 21.89 & 20.75 & 20.21 & 1.13 & 0.54 \\
d & 21.64 & 20.59 & 20.09 & 1.05 & 0.50 \\
e & 21.49 & 20.48 & 19.88 & 1.01 & 0.60 \\
\enddata
\tablecomments{The listed values are corrected for Galactic extinction.}
\end{deluxetable}

\begin{deluxetable}{lr@{.}llr@{.}llr@{.}lr@{.}l}
\tabletypesize{\footnotesize}
\tablecaption{\label{sfr} L$_{\rm H\alpha}$ and SFRs}
\tablewidth{0pt}
\tablehead{
\colhead{Object ID} & \multicolumn{2}{l}{L$_{\rm H\alpha}$}& \colhead{W$_{\rm H\alpha}$}& 
\multicolumn{2}{l}{Q$_{\rm ion}$} & \colhead{N(O5)} &\multicolumn{2}{l}{SFR} & 
\multicolumn{2}{l}{SFRD} \\
 &\multicolumn{2}{r}{(10$^{40}$erg s$^{-1}$)} & (\AA) &
\multicolumn{2}{r}{(10$^{52}$ phot.
s$^{-1}$)} & &\multicolumn{2}{r}{(M$_{\odot}$ yr$^{-1}$)} &
\multicolumn{2}{r}{(10$^{-8}$ M$_{\odot}$ yr$^{-1}$ pc$^{-2}$)}}
\startdata
~3 & 11&56& 198  & 8&44& 1688& 0&82& 14&15\\
~5\tablenotemark{a} &0&06  & ~28  & 0&05& 9& 0&004& 0&05\\
~6\tablenotemark{a} &0&32  & ~24    & 0&23& 46& 0&02& 0&31\\
~7 &5&34  & ~90  & 3&90& 780& 0&38& 3&46\\
~8 &1&30  & ~94  & 0&95& 190& 0&09& 1&41\\
~9 &2&14  & ~26  & 1&56& 313& 0&15& 1&67\\
10 & 1&03 & ~59  & 0&75& 150& 0&07& 0&71\\
\enddata
\tablenotetext{a}{Quoted values for objects 5 and 6 are not corrected for internal extinction because of
the lack of a measurable H$\beta$ line in their spectra.}
\end{deluxetable}

\begin{deluxetable}{cccccc}
\tabletypesize{\footnotesize}
\tablecaption{\label{photoion} Photoionization Model Parameters}
\tablewidth{0pt}
\tablehead{
\colhead{Model Parameter}  & \colhead{Object 3}& \colhead{Object 7}
&\colhead{Object 8} &
\colhead{Object 9}& \colhead{Object 10}}
\startdata
T$_{\ast}$  (10$^4$ K)&4.0  &4.0  & 4.0 & 4.0 &4.0  \\
Z/Z$_{\odot}$         &0.34 &0.30 &0.34 &0.30 &0.10  \\
N/N$_{\odot}$         &0.34 &0.20 &0.34 &0.18 &0.08  \\
O/O$_{\odot}$         &0.34 &0.36 &0.34 &0.36 &0.08  \\
S/S$_{\odot}$         &0.34 &0.36 &0.34 &0.36 &0.10  \\
U (10$^{-4}$)         &4.22 &6.59 &6.93 &4.66 &3.92  \\
N$_{\rm H}$ (10$^2$cm$^{-3}$) &0.94 &1.31 &1.32 &0.57 & 2.27 \\
\enddata
\end{deluxetable}

\begin{deluxetable}{cccccc}
\tabletypesize{\footnotesize}
\tablecaption{\label{obs-mod}Observed and Modeled Line Intensities for the Candidate TDGs}
\tablewidth{0pt}
\tablehead{
\colhead{Line}   & \colhead{Object 3}& \colhead{Object 7}& \colhead{Object 8}& \colhead{Object 9}& 
\colhead{Object 10}}
\startdata
$[$OII$]$ 3727  &\nodata &6.06$\pm$0.19 & 4.37$\pm$0.40 & 6.32$\pm$0.42&2.57$\pm$0.57 \\ 
 &{\bf \nodata } &{\bf 4.70, 2.3E+0} &{\bf 4.16, 1.7E-2} &{\bf 5.11, 3.2E-1}&{\bf 2.64, 2.5E-3} \\ 
H$\gamma$  &\nodata &0.58$\pm$0.38 &\nodata &\nodata & \nodata\\ 
 &{\bf \nodata} &{\bf 0.48, 3.3E-1} &{\bf \nodata} &{\bf \nodata} &{\bf \nodata} \\ 
$[$OIII$]$ 4959  &0.53$\pm$0.89 & 0.83$\pm$0.28&1.01$\pm$0.61 &\nodata & \nodata\\ 
 &{\bf 0.37, 2.4E-1 } &{\bf 0.72, 3.3E-1} &{\bf 0.69, 5.7E-1} &{\bf \nodata} &{\bf\nodata } \\ 
$[$OIII$]$ 5007  &1.00$\pm$0.57 & 2.33$\pm$0.20&1.83$\pm$0.41&1.71$\pm$0.55 &0.59$\pm$0.69 \\ 
 &{\bf 1.07, 1.4E-2 } &{\bf 2.07, 4.1E-1} &{\bf 1.99, 4.8E-2} &{\bf1.35, 2.4E-1} &{\bf 0.60, 1.2E-3} \\ 
$[$NII$]$ 6548 &0.31$\pm$0.89 & 0.17$\pm$1.07&0.24$\pm$1.54 &0.22$\pm$1.68 & \nodata\\ 
 &{\bf 0.43, 1.8E-1} &{\bf 0.20, 2.3E-2} &{\bf 0.35, 9.5E-2} &{\bf0.23, 2.2E-4} &{\bf \nodata} \\ 
H$\alpha$  &2.85$\pm$0.46 & 2.85$\pm$0.19&2.85$\pm$0.40 &2.85$\pm$0.36 & 2.85$\pm$0.50\\ 
 &{\bf 2.86, 2.9E-5} &{\bf 2.85, 7.2E-5} &{\bf 2.85, 1.6E-5} &{\bf2.86, 3.7E-5} &{\bf 2.88, 4.3E-4} \\ 
$[$NII$]$ 6583  &0.71$\pm$0.60 & 0.49$\pm$0.46&0.72$\pm$0.69 &0.64$\pm$0.75 & 0.39$\pm$0.95\\ 
 &{\bf 1.26, 1.7E+0} &{\bf 0.58, 1.7E-1} &{\bf 1.04, 4.2E-1} &{\bf0.67, 2.8E-3} &{\bf 0.50, 9.5E-2} \\ 
$[$SII$]$ 6716  &0.96$\pm$1.00 & 0.66$\pm$0.36&0.62$\pm$0.59 &0.93$\pm$0.55 &\nodata \\ 
 &{\bf 0.62, 3.1E-1} &{\bf 0.53, 4.4E-1} &{\bf 0.48, 2.6E-1} &{\bf0.65, 5.8E-1} &{\bf\nodata } \\ 
$[$SII$]$ 6731 &0.57$\pm$1.3 & 0.57$\pm$0.41&0.53$\pm$0.69 &0.55$\pm$0.64 & \nodata\\ 
 &{\bf 0.46, 3.1E-1} &{\bf 0.41, 8.6E-1} &{\bf 0.37, 4.0E-1} &{\bf0.47, 5.4E-2} &{\bf \nodata} \\ 
\enddata
\tablecomments{Line intensities are relative to H$\beta$. For every emission
line, the observed values with \emph{relative} errors are in the first row and 
the modeled values with the $\chi^2$ of the
fit to the line are in the second row, in boldface type.}
\end{deluxetable}

\begin{deluxetable}{cccc}
\tabletypesize{\footnotesize}
\tablecaption{\label{alt-photoion} Alternative Photoionization Model Parameters}
\tablewidth{0pt}
\tablehead{
\colhead{Model Parameter}  & \colhead{Object 7} &\colhead{Object 8} &\colhead{Object 9}}
\startdata
T$_{\ast}$  (10$^4$ K)&4.0  & 4.0 & 4.0 \\
Z/Z$_{\odot}$         &0.10 &0.15 &0.15 \\
N/N$_{\odot}$         &0.10 &0.23 &0.15 \\
O/O$_{\odot}$         &0.34 &0.38 &0.05 \\
S/S$_{\odot}$         &0.34 &0.38 &0.05 \\
U (10$^{-4}$)         &5.91 &6.45 &4.66 \\
N$_{\rm H}$ (10$^2$cm$^{-3}$) &1.11 &0.95 &0.57 \\
\enddata
\end{deluxetable}

\begin{deluxetable}{ccccccccc}
\tabletypesize{\footnotesize}
\tablecaption{\label{no_bkg} Magnitudes and Colors of TDG Candidates Corrected for Tail Contribution}
\tablewidth{0pt}
\tablehead{
\colhead{Object ID}  & \colhead{$B$} &\colhead{Object $V$} &\colhead{Object $R$} & \colhead{\bv} &
\colhead{\vr} & \colhead{M$_B$} & \colhead{M$_V$} & \colhead{M$_R$} }
\startdata
3+9 & 19.37 & 19.12 & 18.67 & 0.26    & 0.45 & $-$16.9 & $-$17.2 & $-$17.6 \\
7   & 19.87 & 19.81 & 19.33 & 0.06    & 0.47 & $-$16.4 & $-$16.5 & $-$16.9 \\
8   & 23.27 & 23.30 & 22.69 & $-$0.03 & 0.60 & $-$13.0 & $-$13.0 & $-$13.6 \\
11  & 22.54 & 22.87 & 22.26 & $-$0.32 & 0.61 & $-$13.7 & $-$13.4 & $-$14.0 \\
10  & 21.27 & 21.07 & 20.46 & 0.20    & 0.61 & $-$14.2 & $-$14.4 & $-$14.9 \\
12  & 22.05 & 21.84 & 21.42 & 0.21    & 0.42 & $-$15.0 & $-$15.2 & $-$15.8 \\
\enddata
\end{deluxetable}

\end{document}